\documentclass[%
 aip,
 jmp,
 amsmath,amssymb,
]{revtex4-1}

\usepackage{graphicx}% Include figure files
\usepackage{dcolumn}% Align table columns on decimal point
\usepackage{bm}% bold math
\usepackage[utf8]{inputenc}
\usepackage[T1]{fontenc}
\usepackage{mathptmx}
\DeclareMathOperator{\diag}{diag}
\DeclareMathOperator{\Beta}{B}
\newtheorem{prop}{Proposition}
\begin{document}

\title{A general method for rotational averages}
% Force line breaks with \\

\author{R. Nessler}
  \email{rcn@tamu.edu.}
\affiliation{Institute for Quantum Science and Engineering, Texas A\&M University, College Station TX 77840, USA}
% \altaffiliation[Also at ]{Physics Department, Baylor University, Waco TX 76798, USA}
\affiliation{Physics Department, Baylor University, Waco TX 76798, USA}
\author{T. Begzjav}%
\affiliation{Institute for Quantum Science and Engineering, Texas A\&M University, College Station TX 77840, USA}

\date{\today}% It is always \today, today,

\begin{abstract}
The theory of nonlinear spectroscopy on randomly oriented molecules leads to the problem of averaging molecular quantities over the random rotation. We solve this problem for arbitrary tensor rank by deriving a closed-form expression for the rotationally invariant tensor of averaged direction cosine products. From it we obtain some useful new facts about this tensor. Our results serve to speed the inherently lengthy calculations of nonlinear optics.
\end{abstract}

\maketitle

\section{Introduction}
Frequently the theory of nonlinear spectroscopy makes use of tensor descriptions, and the molecules comprising the system have random orientations with respect to the lab-fixed frame. This general situation motivates us to consider uniform rotational averages of $n$th rank three-dimensional tensor quantities. It is well known \cite{Andrews1977} that this problem reduces to calculating a tensor $I^{(n)}_{i_1\cdots i_n;\lambda_1\cdots\lambda_n}$ formed by averaging products of $n$ direction cosines. Writers on the subject oftentimes express this tensor as a trigonometric integral over Euler angles, but evidently not as a productive step in their calculations, for they demur when faced with integrating it. A 2002 textbook \cite{doi:10.1002/3527602747.app2} conveys the attitude:
\begin{quote}
Whilst explicit integration over the Euler angles offers the most obvious means of identifying the [rotational average $I^{(n)}$], it is a method which in principle entails $3^{2n}$ distinct integrals. Despite simplifications that can be effected by exploiting the symmetry properties of the Euler matrix, the procedure remains a formidable task for any $n>2$.
\end{quote}

The desire for a simpler alternative led to the ingenious work of Andrews et al.\ \cite{Andrews1977,Andrews1981} that systematically expresses $I^{(n)}$ in terms of Kronecker and Levi-Civita tensors. Though elegant, these formulas have not been extended past rank $n=8$, where they already involve a very large matrix of coefficients.

It would thus be useful to have a formula for $I^{(n)}$ that is valid for all ranks, fits within a few lines of print or code, and is trivial to run on a computer. Here we provide such a formula, obtained by working out the much-maligned Euler integral representation.

This article takes the following form. We begin by formulating the problem in a way that best suits the calculation and presentation of results. We next consider some symmetry aspects that play an important role in what follows. After that comes the actual calculation of $I^{(n)}$, followed by a brief discussion of applications and conclusion.
\section{Formulation}
By $\mathrm{SO}(3)=\mathrm{SO}(3,\mathbf{R})$ we understand the rotation group of $3\times 3$ real orthogonal matrices of determinant 1. For a pair of indices $i,\lambda\in\{1,2,3\}$ we denote by $l_{i\lambda}\colon\mathrm{SO}(3)\to\mathbf{R}$ the $(i,\lambda)$ coordinate function. We then define
\begin{equation}
\label{haar}
I^{(n)}_{i_1\cdots i_n;\lambda_1\cdots\lambda_n}=\langle l_{i_1\lambda_1}\cdots l_{i_n\lambda_n}\rangle=\int_{\mathrm{SO}(3)}\mathrm{d}g\,l_{i_1\lambda_1}(g)\cdots l_{i_n\lambda_n}(g),
\end{equation}
where $\mathrm{d}g$ is Haar measure on the compact group $\mathrm{SO}(3)$, which has the defining property of left and right invariance. Thus \eqref{haar} is unchanged under $g\mapsto hg$ or $g\mapsto gh$ for any rotation $h\in\mathrm{SO}(3)$, making it the correct notion of ``uniform rotational average''.

By collecting like terms, we may replace $i_1\cdots i_n;\lambda_1\cdots\lambda_n$ by a list of nine powers $Q,R,\ldots,Y$ that sum to $n$:
\begin{equation}
I^{(n)}=\langle l_{11}^Ql_{12}^Rl_{13}^Sl_{21}^Tl_{22}^Ul_{23}^Vl_{31}^Wl_{32}^Xl_{33}^Y\rangle.
\end{equation}
As a mnemonic we arrange the nine powers in a $3\times 3$ array
\begin{equation}
\label{pows}
\chi=\begin{bmatrix}
Q&R&S\\T&U&V\\W&X&Y
\end{bmatrix}
\end{equation}
associating them with the corresponding coordinate functions.

Introducing this notation provides two benefits. First, the number of separate components no longer grows exponentially but polynomially with $n$: $\binom{n+8}8$ instead of $3^{2n}$. Second, as we will see below, the matrix arrangement \eqref{pows} turns out to facilitate some statements.

To realize \eqref{haar} concretely we parametrize $\mathrm{SO}(3)$ using Euler angles. There are multiple ways to assign Euler angles to a rotation but we follow the standard in treatments of angular momentum and the Wigner $D$-matrix \cite{edmonds}, i.e.\ the $z$-$y$-$z$ convention with a right-handed frame of reference and right-handed screw:
\begin{widetext}
\begin{equation}
\label{eulang}
g=\begin{bmatrix}-\sin\alpha\sin\gamma+\cos\alpha\cos\beta\cos\gamma&-\cos\gamma\sin\alpha-\cos\alpha\cos\beta\sin\gamma&\cos\alpha\sin\beta\\
\cos\alpha\sin\gamma+\cos\beta\cos\gamma\sin\alpha&\cos\alpha\cos\gamma-\cos\beta\sin\alpha\sin\gamma&\sin\alpha\sin\beta\\
-\cos\gamma\sin\beta&\sin\beta\sin\gamma&\cos\beta\end{bmatrix}.
\end{equation}
\end{widetext}

The rotational average \eqref{haar} becomes
\begin{equation}
\label{inte}
I^{(n)}=\frac1{8\pi^2}\int_0^{2\pi}\mathrm{d}\alpha\int_0^{\pi}\mathrm{d}\beta\int_0^{2\pi}\mathrm{d}\gamma\,\sin\beta l_{11}^Ql_{12}^Rl_{13}^Sl_{21}^Tl_{22}^Ul_{23}^Vl_{31}^Wl_{32}^Xl_{33}^Y,
\end{equation}
where $l_{i\lambda}$ is given by the $(i,\lambda)$ entry of \eqref{eulang}.
\section{A selection rule}
Before proceeding with the calculation we note a useful consequence of the invariance of \eqref{haar}. If we let $h=\diag(-1,-1,1)$ then invariance under $g\mapsto hg$ implies
\begin{equation}
I^{(n)}=(-1)^{n-(W+X+Y)}I^{(n)}.
\end{equation}
Hence $I^{(n)}=0$ if $n$ and $W+X+Y$ have different parity, i.e.\ one of them is even and the other odd. By varying $h$ we see that $I^{(n)}\neq 0$ only if every row sum of the matrix $\chi$ in \eqref{pows} has the same parity as $n$. Considering instead $g\mapsto gh$ we obtain the analogous statement for columns of $\chi$. In summary, a necessary condition for $I^{(n)}\neq 0$ is
\begin{equation}
\label{sel}
\tag{$\clubsuit$}
\parbox{3in}{The sum along every row and column of $\chi$ is even (if $n$ is even) or odd (if $n$ is odd).}
\end{equation}

We note that \eqref{sel} does not describe a \emph{sufficient} condition, since for instance $I^{(n)}=0$ whenever $n$ is odd and two rows or two columns of $\chi$ are equal. Indeed one can show along the lines above that $I^{(n)}$ for even (resp.\ odd) $n$ is symmetric (resp.\ antisymmetric) in rows and columns of $\chi$. This fact suggests a connection between $I^{(n)}$ and the determinant of $\chi$. That connection, and sufficiency of \eqref{sel} in some but not all even ranks, are discussed in Section~\ref{disc} once the general formula for $I^{(n)}$ is established.

We finally note that $I^{(n)}$ remains the same when transposing $\chi$. This follows from $g\mapsto g^{-1}$ invariance of Haar measure on $\mathrm{SO}(3)$.

\section{Main result}
Our strategy for getting a handle on \eqref{inte} relies on the following integral identities\cite{NIST:DLMF}:
\begin{equation}
\begin{aligned}
\int_0^{\pi}\mathrm{d}x\,\sin^a x\cos^b x&=\frac12(1+(-1)^{-b})\Beta\left(\frac{1+a}2,\frac{1+b}2\right)\\
\int_0^{2\pi}\mathrm{d}x\,\sin^a x\cos^b x&=\frac12(1+(-1)^{-a})(1+(-1)^{-b})\Beta\left(\frac{1+a}2,\frac{1+b}2\right).
\end{aligned}
\end{equation}
Here $\Beta$ denotes the beta function. These are valid whenever $\operatorname{Re}a>-1$ and $\operatorname{Re}b>-1$; in our application $a$ and $b$ are nonnegative integers.

When we expand \eqref{inte} in terms of these integrals we obtain
\begin{widetext}
\begin{equation}
\label{firstmain}
\begin{aligned}
I^{(n)}&=\frac1{64\pi^2}\sum_{q,r,t,u}\binom Qq\binom Rr\binom Tt\binom Uu(-1)^{q+R+U-u+W}(1+(-1)^{c_\beta})(1+(-1)^{c_\alpha})(1+(-1)^{s_\alpha})\\
&\qquad\times(1+(-1)^{c_\gamma})(1+(-1)^{s_\gamma})\Beta\left(\frac{1+c_\alpha}2,\frac{1+s_\alpha}2\right)\Beta\left(\frac{1+c_\beta}2,\frac{1+s_\beta}2\right)\Beta\left(\frac{1+c_\gamma}2,\frac{1+s_\gamma}2\right),
\end{aligned}
\end{equation}
\end{widetext}
where $q$ runs over $0,\ldots,Q$; $r$ runs over $0,\ldots,R$; etc.\ and the collected trigonometric powers are
\begin{equation}
\label{tpows}
\begin{aligned}
c_\beta&=Q+R+T+U+Y-(q+r+t+u)\\
s_\beta&=S+V+W+X+1\\
c_\alpha&=Q+R+S-q-r+t+u\\
s_\alpha&=T+U+V+q+r-t-u\\
c_\gamma&=Q+T+W-q+r-t+u\\
s_\gamma&=R+U+X+q-r+t-u.
\end{aligned}
\end{equation}

Evidently the product
\begin{equation}
(1+(-1)^{c_\alpha})(1+(-1)^{s_\alpha})(1+(-1)^{c_\gamma})(1+(-1)^{s_\gamma})
\end{equation}
vanishes unless $q+r+t+u$ has the same parity as four different sums: those along the first two rows and first two columns of $\chi$. In particular, if those four sums do not all themselves have the same parity then \emph{all} summands of \eqref{firstmain} vanish and $I^{(n)}=0$. This is of course a manifestation of \eqref{sel}.

Let us assume for the remainder of this section that \eqref{sel} holds (for otherwise $I^{(n)}=0$ and a formula for $I^{(n)}$ is not needed). In this case, the sum $Q+R+T+U+Y$ appearing in $c_\beta$ has the same parity as the row sum $W+X+Y$. We can thus replace the factors
\begin{equation}
(1+(-1)^{c_\beta})(1+(-1)^{c_\alpha})(1+(-1)^{s_\alpha})(1+(-1)^{c_\gamma})(1+(-1)^{s_\gamma})
\end{equation}
in \eqref{firstmain} by $2^5$ provided we restrict the sum to include only combinations of indices where $q+r+t+u$ has the same parity as $n$. Moreover, using that $s_\beta$ is odd and the other five numbers listed in \eqref{tpows} are even, we can express the beta functions in terms of double factorials. We obtain
\begin{widetext}
\begin{equation}
I^{(n)}=\sum'_{q,r,t,u}\binom Qq\binom Rr\binom Tt\binom Uu(-1)^{q+R+U-u+W}\frac{(c_\alpha-1)!!(s_\alpha-1)!!(c_\beta-1)!!(s_\beta-1)!!(c_\gamma-1)!!(s_\gamma-1)!!}{(c_\alpha+s_\alpha)!!(c_\beta+s_\beta)!!(c_\gamma+s_\gamma)!!}
\end{equation}
\end{widetext}
where the prime indicates the parity restriction $q+r+t+u\equiv n\pmod{2}$.

Expanding the definitions \eqref{tpows} we arrive at our main result:
\begin{widetext}
\begin{equation}
\label{finalmain}
\begin{aligned}
I^{(n)}&=\frac{(-1)^{R+U+W}\frac{S+V+W+X}2!}{2^{Q+R+T+U}\frac{Q+R+S+T+U+V}2!\frac{Q+T+W+R+U+X}2!}\sum'_{q,r,t,u}\binom Qq\binom Rr\binom Tt\binom Uu(-1)^{q+u}\\
&\qquad\times(Q+R+T+U+Y-q-r-t-u-1)!!(T+U+V+q+r-t-u-1)!!\\
&\qquad\times(Q+R+S-q-r+t+u-1)!!(R+U+X+q-r+t-u-1)!!\\
&\qquad\times\frac{(Q+T+W-q+r-t+u-1)!!}{(n-q-r-t-u+1)!!}.
\end{aligned}
\end{equation}
\end{widetext}

Though \eqref{finalmain} might be tedious to work out by hand with particular powers $Q,\ldots,Y$, it is well suited for evaluation by computer. Its form also has interesting consequences, such as that the components of $I^{(n)}$ are rational numbers, and the less-immediate propositions of the next section.
\section{Discussion}\label{disc}
In practical nonlinear optics calculations, the multiple interacting beams give rise to many terms that need to be evaluated. It is therefore valuable to be able to decide at a glance whether or not a component of $I^{(n)}$ vanishes. The following, obtainable by systematic application of \eqref{finalmain}, settles this question for selected $n$.
\begin{prop}\ 
\label{dets}
\begin{enumerate}
\item[($n$ even)] Suppose $n\in\{0,2,4,6,10,12\}$. Then $I^{(n)}\neq 0$ if and only if \eqref{sel} holds.
\item[($n$ odd)] Suppose $n\in\{1,3,5,7,11,13\}$. Then $I^{(n)}\neq 0$ if and only if \eqref{sel} holds and $\det\chi\neq 0$. Moreover,
\begin{equation}
\begin{aligned}
I^{(3)}&=\frac16\det\chi&\text{in general}\\
I^{(5)}&=\frac1{30}\det\chi&\text{if \eqref{sel} is assumed.}
\end{aligned}
\end{equation}
\end{enumerate}
\end{prop}
Ranks 8 and 9 do not conform to the stated rules, e.g. $I^{(8)}=0$ but \eqref{sel} holds when
\begin{equation}
\label{countereven}
\chi=\begin{bmatrix}
0&0&0\\1&1&2\\1&1&2
\end{bmatrix}
\end{equation}
and $I^{(9)}\neq 0$ but $\det\chi=0$ when
\begin{equation}
\label{counterodd}
\chi=\begin{bmatrix}
1&1&1\\1&2&0\\1&0&2
\end{bmatrix}.
\end{equation}
Up to row and column permutations and transpose, \eqref{countereven} and \eqref{counterodd} provide the only counterexamples in ranks 8 and 9, respectively.

Generalizing \eqref{countereven}, it is easily seen from \eqref{finalmain} that $I^{(n)}=0$ when
\begin{equation}
\label{counterevenhigher}
\chi=\begin{bmatrix}
0&0&0\\1&1&V\\W&VY-W-2&Y
\end{bmatrix}
\end{equation}
where $V\geq 2$ and $Y\geq 2$ are even and $1\leq W\leq VY-3$ is odd. Therefore ranks of the form $n=(V+1)(Y+1)-1$, i.e.\ even $n$ such that $n+1$ is composite, must be excluded from Proposition~\ref{dets}.

For even $n>8$ the matrices \eqref{counterevenhigher} are by no means the only examples where \eqref{sel} holds but $I^{(n)}=0$, even taking symmetries into account. Nonetheless, they do exhaust all even \emph{ranks} where such examples occur. For suppose instead that $n+1\geq 3$ is prime. The $q=r=t=u=0$ summand in \eqref{finalmain} has denominator $(n+1)!!$, while all double factorials in its numerator and in every other summand (if any) are of numbers strictly less than $n+1$. Likewise the binomial coefficients consist of factorials of numbers strictly less than $n+1$. Thus the $q=r=t=u=0$ summand, but no other, contains the prime factor $n+1$ in its denominator when written in lowest terms. It follows that the sum cannot be zero. We have proven the following
\begin{prop}
\label{evprop}
Suppose $n+1$ is an odd prime. Then $I^{(n)}\neq 0$ if and only if \eqref{sel} holds.
\end{prop}

We turn now to odd $n$. The summands in \eqref{finalmain} with $q+r+t+u=1$ total
\begin{equation}
\begin{aligned}
&(Q+R+T+U+Y-2)!!(T+U+V-2)!!(Q+R+S-2)!!\\
&\qquad\times\frac{(R+U+X-2)!!(Q+T+W-2)!![\det\chi-n(QU-RT)]}{n!!}.
\end{aligned}
\end{equation}
Arguing as above we obtain
\begin{prop}
Suppose $n$ is an odd prime. Then $I^{(n)}\neq 0$ if \eqref{sel} holds and $n$ does not divide $\det\chi$.
\end{prop}
The converse is false, however: there are examples in odd prime rank where $\det\chi=0$ or $\det\chi$ is a nonzero multiple of $n$ and $I^{(n)}\neq 0$. We must regard the ``$n$ odd'' part of Proposition~\ref{dets} as an accident of small ranks, unlike the ``$n$ even'' part.

We close by stating some special cases of \eqref{finalmain}, always assuming \eqref{sel}.
\begin{itemize}
\item $Q=R=T=U=0$. If $n$ is odd $I^{(n)}=0$ since the sum is empty. For $n$ even,
\begin{equation}
\label{2x2zero}
I^{(n)}=\frac{\frac{S+V+W+X}2!\frac{S+V+W+X+Y}2!S!V!W!X!Y!}{\frac{S+V}2!\frac{W+X}2!\frac S2!\frac V2!\frac W2!\frac X2!\frac Y2!(S+V+W+X+Y+1)!}.
\end{equation}
Further specializing to $S=W=Y=0$ reveals a $3j$ symbol:
\begin{equation}
I^{(n)}=\begin{pmatrix}\frac V2&\frac X2&\frac{V+X}2\\0&0&0\end{pmatrix}^2.
\end{equation}
We therefore\cite{edmonds} have the identity
\begin{equation}
\langle l_{23}^Vl_{32}^X\rangle=\langle D^{V/2}_{00}D^{X/2}_{00}D^{(V+X)/2}_{00}\rangle
\end{equation}
relating an average of direction cosines to an average of Wigner $D$-matrix elements. Taking instead $S=V=W=0$ in \eqref{2x2zero} produces
\begin{equation}
\langle l_{32}^Xl_{33}^Y\rangle=(-1)^{(X+Y)/2}\langle D^{X/2}_{0\frac X2}D^{Y/2}_{0-\frac Y2}D^{(X+Y)/2}_{0\frac Y2-\frac X2}\rangle.
\end{equation}

\item $Q=1$ and $R=T=U=0$. For odd $n$
\begin{equation}
I^{(n)}=-\frac{\frac{S+V+W+X}2!\frac{S+V+W+X+Y}2!S!V!W!X!Y!}{\frac{S+V+1}2!\frac{W+X+1}2!\frac S2!\frac W2!\frac Y2!\frac {V-1}2!\frac {X-1}2!(S+V+W+X+Y+1)!}
\end{equation}
and in particular
\begin{equation}
\langle l_{11}l_{23}^Vl_{32}^X\rangle=-\langle D^{V/2}_{\frac 12\frac 12}D^{X/2}_{-\frac 12-\frac 12}D^{(V+X)/2}_{00}\rangle.
\end{equation}

For even $n$
\begin{equation}
I^{(n)}=-2\frac{\frac{S+V+W+X}2!\frac{S+V+W+X+Y+1}2!S!V!W!X!Y!}{\frac{S+V+1}2!\frac{W+X+1}2!\frac {S-1}2!\frac {W-1}2!\frac {Y-1}2!\frac V2!\frac X2!(S+V+W+X+Y+2)!}.
\end{equation}
\end{itemize}

\section{Conclusion}
We have derived a formula for three-dimensional rotational averages of direction cosine products that holds in complete generality, with no restriction on rank, thereby clearing the path for any three-dimensional cartesian tensor to be averaged. The formula is expressed using simple arithmetic---avoiding the ever-larger matrices of earlier methods---making it not only easy to compute but also a friendly base for deriving further results. As illustration we obtained simple criteria to determine when $I^{(n)}=0$ and drew a connection to Wigner's $D$-matrix. We hope that future work will extend these results and uncover additional applications.

\begin{acknowledgments}
We wish to acknowledge G.\ Agarwal for useful discussions. We acknowledge the support of Office of Naval Research Award No.\ N00014-16-1-3054 and Robert A.\ Welch Foundation Grant No.\ A-1261.
\end{acknowledgments}

\nocite{*}
\bibliography{rots.bib}% Produces the bibliography via BibTeX.

\end{document}